\documentclass[conference]{IEEEtran}
\IEEEoverridecommandlockouts
\usepackage{cite}
\usepackage{amsmath,amssymb,amsfonts}
\usepackage{algorithmic}
\usepackage{graphicx}
\usepackage{float}
\usepackage{textcomp}
\usepackage{xcolor}
\usepackage{multirow}
\graphicspath{ {./images/} }
\def\BibTeX{{\rm B\kern-.05em{\sc i\kern-.025em b}\kern-.08em
    T\kern-.1667em\lower.7ex\hbox{E}\kern-.125emX}}
\begin{document}

\title{UniNet: Next Term Course Recommendation using Deep Learning\\
\thanks{Identify applicable funding agency here. If none, delete this.}
}

\author{\IEEEauthorblockN{Nicolas Araque Volk}
\IEEEauthorblockA{\textit{Faculty of Engineering} \\
\textit{Universidad Metropolitana}\\
Caracas, Venezuela \\
naraque@unimet.edu.ve}
\and
\IEEEauthorblockN{Germano Rojas}
\IEEEauthorblockA{\textit{Faculty of Engineering} \\
\textit{Universidad Metropolitana}\\
Caracas, Venezuela \\
germano.rojas@correo.unimet.edu.ve}
\and
\IEEEauthorblockN{Maria Virginia Vitali}
\IEEEauthorblockA{\textit{Faculty of Engineering} \\
\textit{Universidad Metropolitana}\\
Caracas, Venezuela \\
maria.vitali@correo.unimet.edu.ve}

}

\maketitle

\begin{abstract}
Course enrollment recommendation is a relevant task that helps university students decide what is the best combination of courses to enroll in the next term. In particular,  recommender system techniques like matrix factorization and collaborative filtering have been developed to try to solve this problem. As these techniques fail to represent the time-dependent nature of academic performance datasets we propose a deep learning approach using recurrent neural networks that aims to better represent how chronological order of course grades affects the probability of success. We have shown that it is possible to obtain a performance of 81.10\% on AUC metric using only grade information and that it is possible to develop a recommender system with academic student performance prediction. This is shown to be meaningful across different student GPA levels and course difficulties. 
\end{abstract}


\section{Introduction}

The curriculum of the system engineering major from the Universidad Metropolitana helps students understand what are the expected academic requirements in order to graduate from that major. Even when this curriculum states all the required courses and gives a recommendation on what to enroll on a term basis, there is also a lot of flexibility. This flexibility gives students the possibility to plan their itinerary more personally, taking into account schedule, time commitment, etc. This curriculum flexibility is a common approach for numerous universities in the world. 

But this flexibility presents university students with a difficult decision when they complete every academic term: how many courses should I enroll in the next term to keep up to date with the curriculum and also to maintain the desired GPA?. And for every course that they choose, another decision comes to mind: is this course combination is going to be too difficult to carry on?

Universities have been able to provide some solutions to this recurrent problem, with study plans that recommend how courses should be taken and in what order. But this solution is a one fit for all that doesn't do well when some difficulties are encountered along the way. Expert advice from counselors is a great solution that is not scalable to thousands of students within an in-campus university setting. This lack of guidance is a real issue for student success and it has been found that supporting students is key for preventing student dropout [2].

There has been research [3] that tries to predict which students need support in order to intervene and to prevent dropout. This strategy does not take into account that all students can benefit from the help and expert advice, from the high probability dropout student who needs help to stay in college to the low probability dropout student who needs help improving the GPA. 

We could find works that focus on recommender system techniques such as matrix factorization methods [5] and decision trees [13] but these approaches failed to represent the data as a time series, where students take courses along several years of their academic life.  

We propose a deep learning recommender system based on recurrent neural networks that can help every student to make better decisions about how many courses and in what combinations it is appropriate to enroll in order to obtain better academic results. Our system relies only on the academic performance of previous academic terms to predict the probability of passing all credits for the next period given a combination of courses to be registered. 

We defined as academic success the event of obtaining a passing grade in all the enrolled courses for a given term. This definition of academic success is related to the successful fulfillment of the academic itinerary for the student to finish the career.

The main contributions this paper presents are: 

\begin{itemize}
\item The data pre-processing pipeline for conserving time-dependent features of grade transcription data.
\item The use of Deep Learning for course recommendation in an on-campus university setting using Long Short Term Memory recurrent neural network.
\end{itemize}

In the remainder of this paper, we discuss some related work on student performance prediction and recommendation for course enrolment in section 2, Section 3. Details the data pre-processing step where we build the pipeline for a time-dependent input vector followed by an explanation of the UniNet architecture in Section 4, with the analysis of results in Section 5 and conclusions and future work in Section 6.  

\begin{figure}[h]
    \centering
    \includegraphics[width=0.5\textwidth]{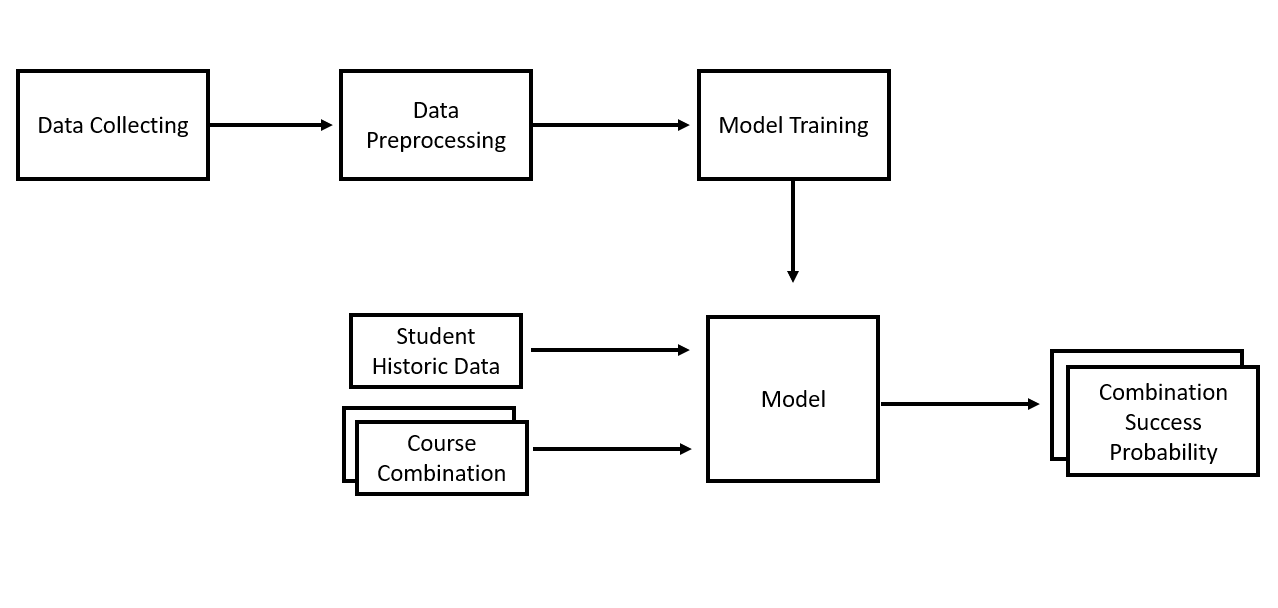}
    \caption{ Main steps for the proposed model}
    \label{fig:my_label}
\end{figure}

\section{Related Work}

Student performance prediction is a large research topic in the  educational data mining area. From machine learning methods like decision trees, k-nearest neighbor, rule-based, SVM, PCA, and logistic regression [11][1][10] to deep learning methods like artificial neural networks and recurrent neural network [7][4][10].

There are also numerous researches using recommender system techniques like matrix factorization and collaborative filtering [2] [6]. This approach fails to represents the time dependent features of the data, and often leads to the neccesity of making one model per term or per year [5].

A lot of student performance prediction research is focused on online educational settings called MOOCs, using Udacity [7], Coursera, and other online platform datasets [4]. In online educational settings research tends to design models that focus short-term data like quizzes and homework within a course [5].

Recurrent neural networks RNN are deep learning architectures that use a recurrent connection to model sequential and time-series data [14]. These recurrent connections enable the model to learn, store, and process information from long time periods. The use of recurrent neural networks for student performance prediction is an area that has not been worked broadly. Moreover, the use of this deep learning architecture for course recommendation and planning is a greenfield for study [4].

 Specifically, long-short term memory LSTM is an effective and scalable type of RNN [15] used extensively to predict time-ordered data where the distance in time from observations needs to be taken into account for the prediction [9]. It has been studied that recurrent neural networks show an improvement in performance when compared with artificial neural networks for student prediction task [12]. We choose this type of architecture to model the possibility that grades from early terms in time are less relevant than grades from recent ones for the next term prediction.

Using LSTM as a recommender system techniques for time-dependent data has reported good performance when compared to traditional techniques like nearest neigbords and matrix factorization [8]. 

\begin{figure*}[t]
    \centering
    \includegraphics[width= \textwidth]{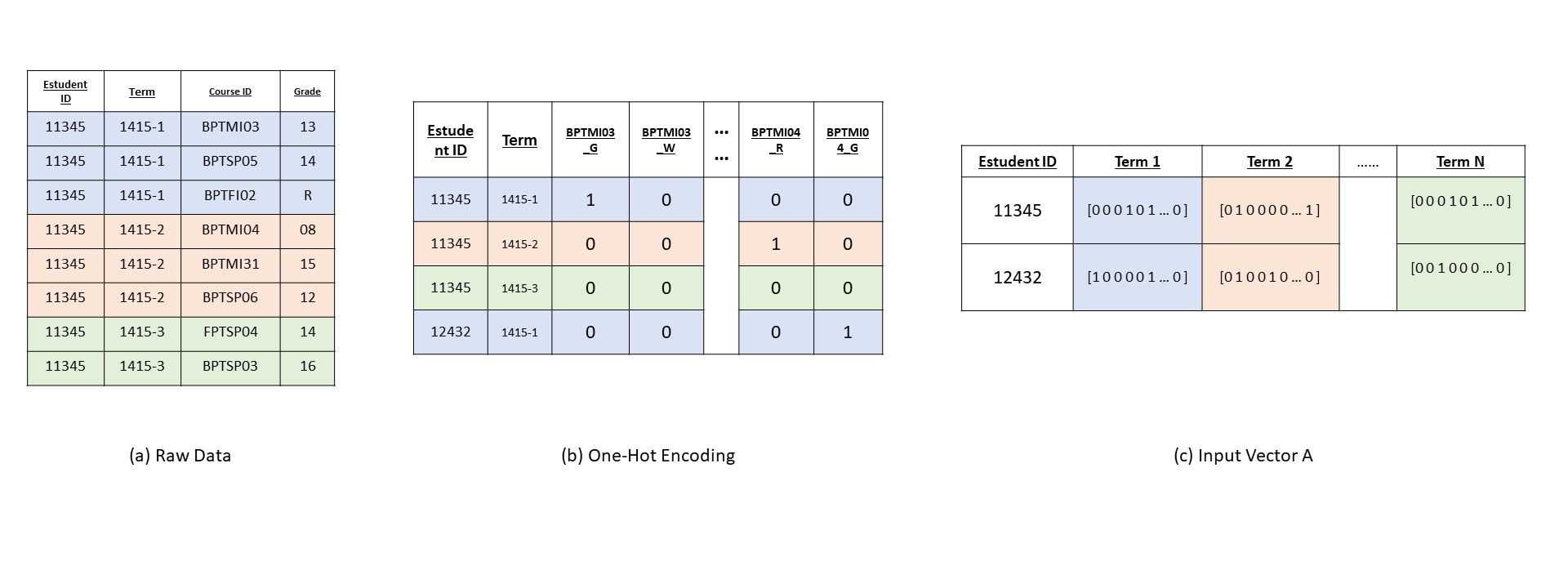}
    \caption{The data preprocessing pipeline with the following steps: (a) the raw data that comes from the university systems, (b) one-hot encoding of all courses for a student in one term, and (c) input vector of all student history ordered chronologically. }
    \label{fig:my_label}
\end{figure*}

\section{Data pre-processing}

Figure 1  shows the main process followed in this paper. First, we extract raw academic data from the university systems. This data is preprocessed and fed to the model for training. The resultant model is then made available for general use. 

Raw academic data comes from the university student life cycle system as it is show on figure 2. This information contains the academic performance of system engineering university students from 2010 to 2018 where each row represents a grade from a student in a particular course in a particular period. Our database has 23719 students, course, period combinations with 817 distinct students. The grading system is numerical and is represented by numbers from 0 to 20 where 10 is the pass mark. There is also the possibility that the student withdraws the course before the end of the academic term, and this is represented with an 'R' in the original dataset. 

To low dimensionality, we bucket the numerical grades to 4 categories: withdraws, not approved (less than 10), bad (from 10 to 12),  and excellent (more than 12). The bad category is included because even when the passing grade is more than 10 points one of the university policies states that a GPA greater than 12 is needed for graduation. From this, we created a multi label one-hot encoding representing a single academic term with all possible course and grade categories combination with a 1 on those courses and categories the student got that term as shown in figure 2. With this step the numbers of rows went from 23719 (for student, period, course) to 5351 (student, period, all courses that where taken that term). This multi label one-hot encoding represents a time step for our recurrent neural network model.

The last step is to create a single input vector for every student to the model. The terms are arranged in chronological order from left to rights. The last period of every student is removed from this vector and becomes the ground truth. If all courses were passed on the last term, the ground truth is going to have a value of 1 and, if not, it is going to be 0.

This final dataset has 817 rows that represent the complete academic history of our student's sample and is split with 773 rows used for training, and the remaining 44 are used for validation.

\section{Recommendation Using Deep Learning}

The objective of the model is to predict what is the probability some particular student has of passing all of the courses he wants to enroll in the next term.  The student can query the model with several plans to compare which one has more probability of success and therefore which course combination is better to enroll. 

Figure 3 shows the model architecture with the inputs and outputs. The model takes two inputs to make a prediction, the individual student's academic performance that is represented with the input vector A, and the course combination for which the success prediction is going to be performed represented with the input vector B. The output is going to be a probability of success for each of these combinations.

The input vector A aims to represent the past academic history of the student in chronological order. This representation enables the LSTM to model the importance not only that previous courses has on futures ones, but also how the time that has passed since that course affects this correlation. Intuitively this feature is going to mimic how students learn and forget information in real life using the input, output and forget gate from the LSTM. 

The input vector B represents the course combination for which the student wants to get the prediction. This representation models the idea that the probability of success for a given term is defined by the combination of courses that the students want to enroll and not by the union of the independent probabilities of success for each course. This feature aims to represent how some courses are designed to be taken together while others are recommended not to.

The output represents the academic success probability that this particular student has with this particular course combination.  For a single course combination, this probability is hard to understand for students because the quantity is not self-explanatory. For the student, it is not sufficient to make a decision this single number representing a probability for a single course combination. 

The use of the model makes sense when queried for several course combinations, so the student can take decisions based on the path with a higher probability of success regardless of the true meaning of this number.

\begin{figure}[h]
    \centering
    \includegraphics[width=0.5\textwidth]{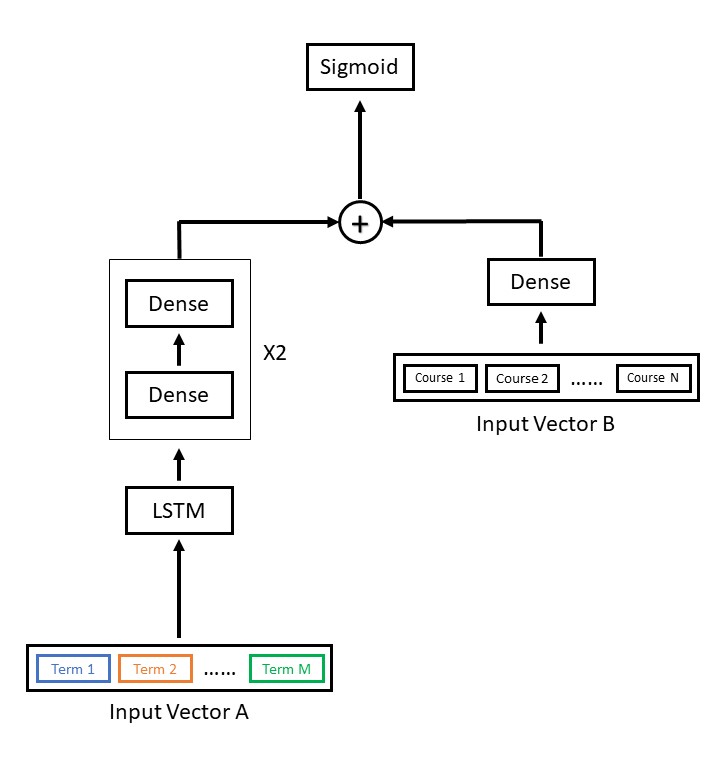}
    \caption{Model Architecture where the left side inputs the student academic data (input vector A) and the right side inputs the course combination to be queried (input vector B)}
    \label{fig:my_label}
\end{figure}

\section{Results}

Using this model we were able to obtain 81.10\% performance using the Area Under the Curve metric for the validation set. The AUC value for the training set is 86.16\%. 

In order to understand the quality of the recommendation beyond accuracy metrics we carried out an experiment to understand how the model predict different term combination for students with different GPA. For this experiment, we selected 3 different term combinations with different historic difficulty. The first term combination contains 80\% of courses with a failure rate over 30\% and is considered a difficult term to succeed. Term combination 2 and 3 has 50\% and 0\% courses with a failure rate over 30\% and are considered medium and easy  terms to succeed.   

Then we selected several students from the validation set with different historical performances (GPA below 12, GPA below 16 and over 16) to understand how different combinations of difficulty from the next term changed the output of the model for different students. As shown in figure 4, the model performed as expected by the System Engineering School director. Students with high GPA have an overall better probability of success that students with medium and low GPA, but the difficulty of the course combination does decrease this probability for all type of students. 

\begin{figure}[h]
    \centering
    \includegraphics[width=0.5\textwidth]{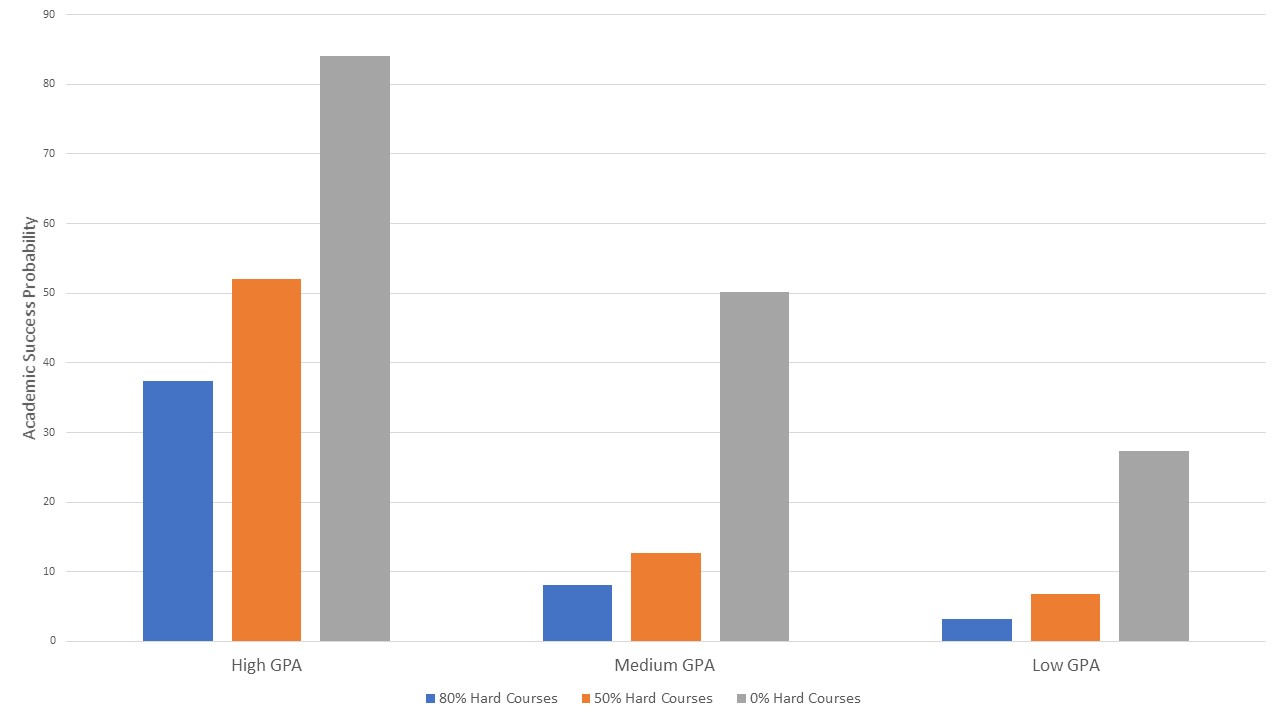}
    \caption{Model output: Probability of success for students with different GPA over courses combinations with decreasing difficulty.}
    \label{fig:my_label}
\end{figure}

\section{Conclusion and future work}

In this paper, we have successfully applied a recurrent neural network for academic performance prediction and used it as a recommendation technique for next term course combination. In contrast to prior work on this task, we have only used past academic performance (course grades) ordered chronologically to predict next term academic success.

The use of bidirectional LSTM was a key component of the architecture to model how grades evolve from the first term of the students to later ones and how this evolution influences the probability of success in the future terms. 

Two novel properties of UniNet are that use chronological information from past terms as a feature to better predict student performance for the next period. Our model also has the ability to recommend not course wise (giving a recommendation on a specific course) but widely on course combination. This is an ideal feature because the overall academic success for the predicted term is not only dictated by the complexity of each individual course enrolled but also, the complexity of the combination of courses as a whole

One main obstacle for scalable deployment of this model is the size of the input vector, which needs to represent all possible courses multiplied by all possible categories that these courses can be (withdraw, not pass, bad and good). In future work, we are aiming to develop a dense input vector representation for the historic academic terms with their respective grades.

\end{document}